# Genomic Analysis of Date Palm Fruit Size Traits and Identification of Candidate Genes through GWAS

Shameem Younuskunju[1,5], Yasmin A. Mohamoud[1], Lisa Sara Mathew[4], Klaus F. X. Mayer[5,6], Karsten Suhre[3] and Joel A. Malek[1,2]*

[1]Genomics Laboratory, Weill Cornell Medicine-Qatar, Doha, 24144, Qatar.

[2]Department of Genetic Medicine, Weill Cornell Medicine-Qatar, Doha, 24144, Qatar.

[3]Department of Physiology, Weill Cornell Medicine-Qatar, Doha, 24144, Qatar.

[4]Clinical Genomics Laboratory, Sidra Medicine, P.O Box 26999, Doha, Qatar.

[5]Shool of Life Sciences, Technical University of Munich, 85354, Munich, Germany.

[6]Plant Genome and Systems Biology, Helmholtz Center Munich, 85764, Munich, Germany

*Corresponding Author:
Joel A. Malek
e-mail: jom2042@qatar-med.cornell.edu
tel: +974-4492-8420

## Abstract

The commercial value of economically significant fruits, including date palm fruit (dates), is influenced by various factors, such as biochemical composition and morphological features like size, shape, and visual appearance, which are key determinants of their quality and market value. Dates are typically consumed at the dry stage (Tamar), during which they exhibit a wide range of physical characteristics, such as color, length, weight, and skin appearance. Understanding the genetic basis of these traits is crucial for improving crop quality and breeding new cultivars. In this study, we integrated a genome dataset from highly diverse date cultivars with phenotypes of dry fruit such as length, width, area, and weight, identifying multiple significant genetic loci (SNPs) associated with these traits. We also identified candidate genes located near the associated SNPs that are involved in biological processes such as cell differentiation, proliferation, growth, and the regulation of signalling pathways for growth regulators like auxin and abscisic acid, as observed in other plants. Gene expression analysis reveals that many of these genes are highly expressed in the early stage of fruit development when the fruit attains its maximum size and weight. These findings will enhance our understanding of genetic determinants of fruit size particularly at the commercially important Tamar stage.

## Keywords

## 1. Introduction

The commercial value of economically significant fruits, including date palm fruit (dates), is determined by the fruit's biochemical composition and morphological features such as color, size, shape and absence of skin defects. Date palm (Phoenix dactylifera L.) is an important crop in regions like the Arabian Peninsula, North Africa, and South Asia, where it contributes significantly to the local economy. Dates are a rich source of nutrition and are known as the 'Bread of the Desert' due to the long-term productive nature of the tree [1, 2]. They are typically consumed at the dry stage due to their low tannin content, long shelf life, high sugar levels, and low moisture [3, 4]. Different date cultivars produce fruits with varying sizes, shapes, skin textures, and colors, which are key parameters for assessing their quality and market value [5-7]. The growth phase of dates involves many developmental stages, such as Hababauk, Kimri, Khalal, Rutab, and Tamar [8]. At the Khalal stage, the fruit becomes physiologically mature and gains maximum weight and size (length and weight).

Recently, interest in studying the genetic basis of traits in date palms has grown, partly due to their economic significance and the development of necessary genetic resources [5, 6, 9-13]. There have been many genetic basis studies on the physical characteristics and biochemical composition of dates, including ours [5, 6, 9-14]. In our previous association study of colour variation of dry date plam, we revealed that there are numerous genetic markers associated with the lightness of the brown shade in dried fruit in addition to the previously identified VIR genotypes associated with the red and yellow colour of fresh dates [6, 12]. In the study of the genetic basis of the skin separation phenomenon, we demonstrated that both environmental and genetic factors contribute to the fruit skin separation of dates, and genetic factors play a dominant role in cases of extreme skin separation observed in certain cultivars [5]. A genome-wide association study of phenotypes of fresh dates by Hazzouri et al. assessed traits such as length and weight [12]. However, they could not map the genetic association with the length and weight phenotype, which might be due to less diversity in the population used for the study.

The length of a fruit is affected by the elongation, and it includes biological mechanisms, such as cell proliferation, differentiation, and expansion [15, 16]. Numerous studies have been conducted to understand the critical regulatory elements and genes associated with the growth and development of other fruits [17-21]. Tomatoes, cucumbers, and peaches have been widely studied to understand the regulatory elements and genes associated with fruit shape, and research has identified several critical regulatory elements and genes associated with fruit development [17-19, 22]. In tomatoes, the fruit's shape and size are influenced by both hormone regulation and gene pathways [23, 24] . The genes CLV, LC, SUN and OVATE contribute to fruit elongation (length) and shape in tomatoes and chilli peppers [20, 21, 24, 25]. In Arabidopsis, the cell elongation and development are controlled by the OFP member, which acts as a transcriptional repressor [26].

Understanding the genetic architecture underlying the physical characteristics of dates, especially at the dry stage (Tamar), is essential for improving crop quality and developing new cultivars with desirable traits. In this study, we focused on identifying the significantly associated genetic loci and possible candidate genes linked to the variation in fruit size phenotypes of Tamar stage dates. We utilized a genetic dataset comprising highly diverse samples from our previous GWAS studies [5, 6]. This dataset includes both genetically similar and different cultivars grown in various environments. To our knowledge, no prior studies have explored the influence of genetic factors on the phenotypes of dry date fruit, such as length, width, area, and weight.

## 2. Materials and Methods

**Fruit size Phenotype: length, width, area and weight**

We used fruit photographs of 179 cultivars from our previous association studies to measure the phenotype such as fruit length, width, area and weight [5, 6]. Each image represents a cultivar with 5 to 11 fruits from the same collection as representatives (Supplementary file: Figure. S1). For detailed information on images, please refer to the study by Younuskunju et al [6]. We measured the calliper diameter (length), minimum diameter (width), and fruit area from fruit images using MIPAR software [27]. Fruits were manually assessed, and weight (g) was measured at the time of fruit collection. Multiple images of each cultivar were measured separately, and then averages of each phenotype for each cultivar were calculated. Outliers were removed from the raw phenotype dataset using the Z-score method ( with ± 2 standard deviations), and the mean score for each cultivar was subsequently calculated. This dataset is used as a phenotype for the association study.

**Genome Alignment and SNP calling**

Genome data sets of 199 cultivars from our previous association study on the date fruit color were used for this study [6]. Quality control (QC) processing of samples and raw reads, genome alignment, and variant calling were conducted following the methodology of our previous association study. SNPs were marked as missing if depth (DP) was below 10. Further filtering was performed using VCftools (v0.1.16) with the following criteria, genotype call rate 80%, minor allele frequency 0.01, and Hardy-Weinberg equilibrium $1\times10^{-6}$. The filtered samples and SNPs were then used for subsequent analysis.

**Genome-wide Association study**

Genome-wide analysis (GWAS) was performed on the LD pruned dataset using the FarmCPU method [28] implemented in the GAPIT (v3) R package. We executed LD pruning with PLINK software [29] using the '--indep-pairwise 500 50 0.99' option to improve the computational efficiency of the FarmCPU method. A kinship matrix and four principal components (PCA) were used as covariates in the GWAS to correct the population structure. Both the PCA and kinship matrix scores were calculated from the LD-pruned SNPs using the GAPIT R package. The VanRaden algorithm in the GAPIT R package was used to measure the kinship matrix. A list of significant SNPs associated with phenotype was identified using a cutoff value of FDR-adjusted p-values of 0.05 (5%). Manhatten and QQ plots were generated from association results using the CMPlot R package [30].

**Gene expression and structural variation study**

The 100kb region around the GWAS-identified significant SNPs was considered a potential region for candidate genes and variants. Sequences of possible candidate genes were mapped from the PDK50 reference genome (PRJNA40349) using the GFF3 annotation file. Gene ontology and function analysis of genes were performed using Blast2GO software and literature reviews [31]. All SNPs and INDEL from the potential region were annotated using the SNPEff software and then filtered with an LD R2 value >=0.6 to the significantly associated SNPs [32]. Gene expression analysis was performed using the transcriptome data of two cultivar data (kheneizi and Khalas ) from Hazzouri et al.'s study [12]. The data contain three or four replicates taken at different post-pollination days (DPP) in two cultivars (45,75,105,120,135 days). Read alignment and expression analysis was performed as described in our previous association study on fruit color . Structural variation analysis (SV) was conducted by utilising clipped, discordant, unmapped, and indel reads from each sample that has significantly associated SNPs with a homozygous allele alternative (ALT) to the reference genome. These reads were grouped based on ALT alleles, and structural variation analysis was performed. Refer to the study by Younuskunju et al. for details on the analysis of SV [6].

## 3. Results

**Phenotypic data analysis: Fruit size measurement**

For the association study, we phenotyped the fruit size traits of dry-stage fruit (Tamar), including length, width, area and weight. The size measurements analysis revealed significant variation among cultivars (Supplementary file: Figure. S1). The average fruit length across 179 cultivars ranged from 2.5 cm to 6.1 cm, while the width varied between 1.9 cm and 2.5 cm (Additional file 1). Fruit weight spanned from 3 g to 25 g. The analysis shows that cultivars such as Deglet, Medjool, Amber-medina, Ambara, and Nabt produced larger-sized fruits, whereas Dabbas, Tantebouchte, Ayash, Ajwa medina Msalla exhibited smaller-sized fruits (Figure 1).

**Variant calling and Association study**

Data quality control (QC) and linkage disequilibrium (LD) pruning were conducted as described in our previous association study on fruit color [6], and subsequently, we performed the GWAS of fruit size phenotypes. QC filtering of the raw genotypic dataset resulted in 188 samples and 10,183,993 SNPs across 18 linkage groups (LGs) and unplaced scaffolds in the reference genome. After LD pruning, 3.541 million SNPs were retained for the association study. For detailed information on SNP calling and the association study, please refer to the study by Younuskunju et al [6]. The GWAS of length, width, and weight phenotypes identified several significant associations. An FDR-adjusted p-value cutoff of 5% ( FDR < 0.05) identified five significant SNPs associated with the length phenotype, four with width and one with the weight phenotype (Figure 2 and Table 1). These SNPs were located across multiple LGs in the PDK50 reference genome. Analysis of the phenotypes across the three possible genotypes at these SNPs indicates that the differences in fruit size are linked to genotypic variation (Figure 3). Among the

GWAS-identified SNPs associated with length phenotype, SNPs LG3s41201927, LG4s45025843, LG4s76101205, and MU009046.1s78932 showed significant Wilcoxon test p-values (Figure 3a). Additionally, all SNPs associated with width and weight phenotypes also showed significant associations (Figure 3b & 3c). Manual analysis of fruit images, along with the genotype of SNP LG4s45025843 (from the GWAS results for the length phenotype), confirmed this finding. It showed that the cultivar produces longer fruit when the sample is either homozygous for the alternate allele (ALT) or heterozygous (HET), while smaller fruit is produced when the sample is homozygous for the reference allele (REF) (Figure 4). The association study of the fruit area phenotype revealed no significant associations (Supplementary Figure S3), suggesting that larger populations may be needed to identify the genetic factors underlying this trait.

**Candidate gene and expression and structural variation study**

A total of 186 genes were identified across all potential regions of significant SNPs based on the GWAS results (Additional file 3). Blast2Go analysis and literature searches revealed that many of these genes are involved in developmental processes, response to ethylene, auxin-activated signalling pathways, regulation of unidimensional cell growth and growth rates in other plants (Table 2). Gene expression analysis using transcriptomic data showed that many of these genes were expressed at various stages of fruit development (Figure 5). The SNPEff annotated SNPs and INDELs from the candidate regions with an LD R2 value >=0.6 to the significantly associated SNPs showed many putative high and moderate effects on encoded proteins (Additional file 4). Structural variation (SV) analysis of regions associated with significant SNPs in the length, width, and weight phenotypes revealed various structural variations, including insertions, deletions, and repeat expansions (Additional file 5). However, none of these SVs demonstrated a strong correlation with the genotypic variation of the significant SNPs identified in the GWAS findings.

## 4. Discussion

Domesticated crops exhibit significant genotypic and phenotypic variation, including changes in fruit size, color, weight, and seed dispersal compared to their wild relatives (known as domestication syndrome) [33-36]. This variation is primarily driven by strong artificial selection. Many research studies have shown that a small number of genes and loci may be involved in the domestication syndrome [33, 37]. Identifying and understanding the genetic architecture of these genes is crucial for improving fruit quality and potentially developing new cultivars with desirable traits, especially for economically important crops, including date palms. Dates are most often consumed at the dry stage (Tamar). At this stage, they exhibit a wide range of physical characteristics, such as colour, length, weight, and skin appearance [3, 5, 6], which are key in determining their quality and market value. Our size measurements analysis revealed fruit size varied from cultivar to cultivar, and they ranged from small fruit (length 2.5cm) to bigger size (length 6.1cm) (Supplementary file: Figure. S1). By integrating a genome data set of highly diverse date cultivars with dry fruit traits such as length, width and weight, we identified multiple significant genetic loci (SNPs) associated with these phenotypes. These SNPs are present across multiple linkage groups (LG) of the reference genome. The phenotype by genotype analysis of the GWAS-identified SNPs revealed three possible genotypes at the GWAS-identified SNPs that are linked with the variations in fruit size phenotypes (Figure 3). The analysis results for SNPs from GWAS result of length phenotype, LG3s41201927, LG4s45025843, LG4s76101205, indicated that fruit length increases when the sample is either homozygous alternate allele (ALT) or heterozygous (HET) for the reference allele (Figure 3a). Similarly, an increase in fruit width was observed when the sample was homozygous ALT or HET for SNPs LG4s20658859, LG2s55487560, LG4s77755700, and LG10s25433217 (Figure 3b). Additionally, the analysis of SNP LG4s73547814 demonstrated that fruit weight increases when the sample is homozygous ALT or HET for the reference allele (Figure 3c). The overall picture is that the GWAS-identified loci are associated with the fruit size phenotype variation and may relate to genetic control during the developmental stages of fruit.

In addition to the identification of significant SNPs associated with fruit phenotypes, we identified several potential candidate genes that are present within the regions of GWAS-identified

SNPs. Genes such as Protein kinase superfamily protein, Myb transcription factor, Mitogen-activated protein kinase, 3-ketoacyl-CoA synthase, and Auxin efflux carrier component were found in the regions associated with the length phenotype (Table 2). These genes are known to be involved in biological processes, such as cell differentiation, proliferation, unidimensional cell growth, and regulation of signalling pathways of growth regulators such as auxin and abscisic acid in other plants [16, 19, 24, 38-41]. Additionally, Growth-regulating factor 5, Cytochrome P450, and GPCR-type G protein-2 genes were found in the regions associated with the width phenotype. 3-ketoacyl-CoA synthase is a gene-encoding enzyme involved in the synthesis of very long-chain fatty acids (VLCFAs) in plants, which are essential for plant growth [42-44]. A study by Nobusawa et al. 's study showed that the VLCFAs synthesised in the epidermis are required to suppress cytokinin biosynthesis, thereby controlling cell overproliferation and ensuring proper organ growth in *Arabidopsis* [42]. Growth-regulating factor5 (GRF5) is a plant-specific transcription factor involved in the regulation of cell proliferation and thus plays a key role in growth [45-48]. Auxin is a crucial regulator of plant development, with its accumulation patterns directing the initiation and growth of plant organs, such as leaves, roots, and flowers [49-52]. Auxin efflux carriers play a significant role in exporting auxin from its synthesis sites in young apical parts and facilitating its accumulation in target tissues, thereby regulating auxin translocation and contributing to plant development [53-55]. The growth and development of dates involve five distinct stages, namely Hababauk, Kimri, Khalal, Rutab, and Tamar, with the fruits reaching physiological maturity and gaining their maximum weight and size at the Khalal stage [56]. Our gene expression analysis revealed that genes like 3-ketoacyl-CoA synthase, Auxin efflux carrier component, and Growth-regulating factor 5 were highly expressed during the early stages of fruit development, particularly in the Hababuk, Kimri and Khalal stages (Figure 5). The expression patterns of these genes, along with the anatomical changes observed in the fruit during these developmental stages, can correlate with each other. This result indicates that these genes may be important for the size regulation of dates. However, further research is needed to determine their specific functional roles in phenotype variation.

By combining genotypic data with fruit size phenotypes (length, width, and weight), we successfully conducted an association study and mapped several significant genetic loci associated with these traits. The identified candidate regions and genes offer valuable opportunities for further functional studies, not only in dates but also in other fruit crops.

## 5. Figures and Table

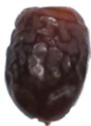

**Figure 1**: Phenotype comparison analysis of date cultivars with the lowest and highest phenotype values. Cultivars such as Deglet, Medjool, Amber-medina, Ambara , and Nabt produced larger-sized fruits, whereas Dabbas, Tantebouchte, Ayash, Ajwa medina Msalla showed smaller-sized fruits.

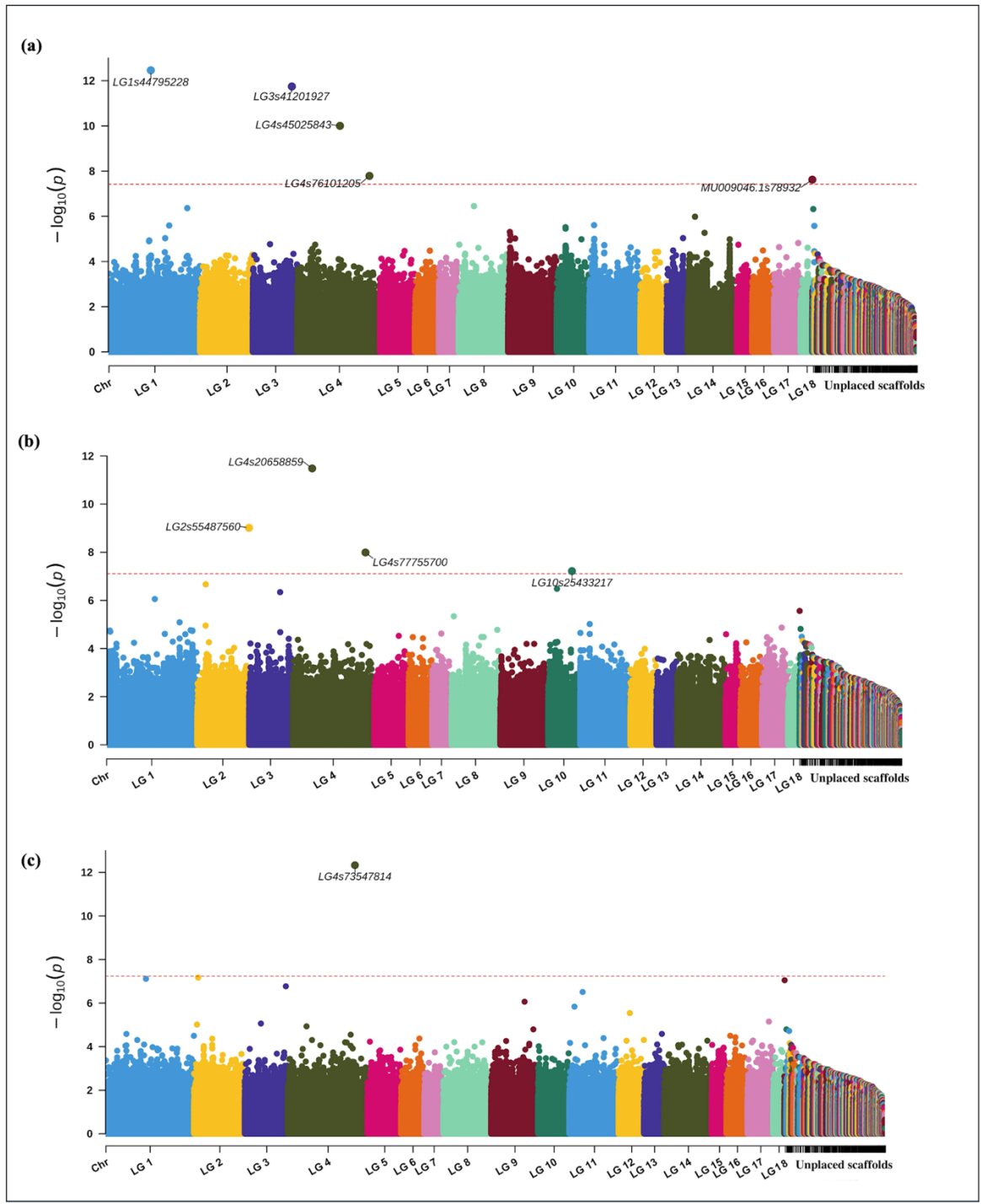

**Figure 2**: Genome-wide association study (GWAS) analysis results for the length, width, weight, and area phenotypes using the LD-pruned SNP set. **(a, b, c)** : Manhattan plots for length, width and weight phenotypes of dates ,respectively, using the LD-pruned SNP set across all linkage groups and unplaced scaffolds.

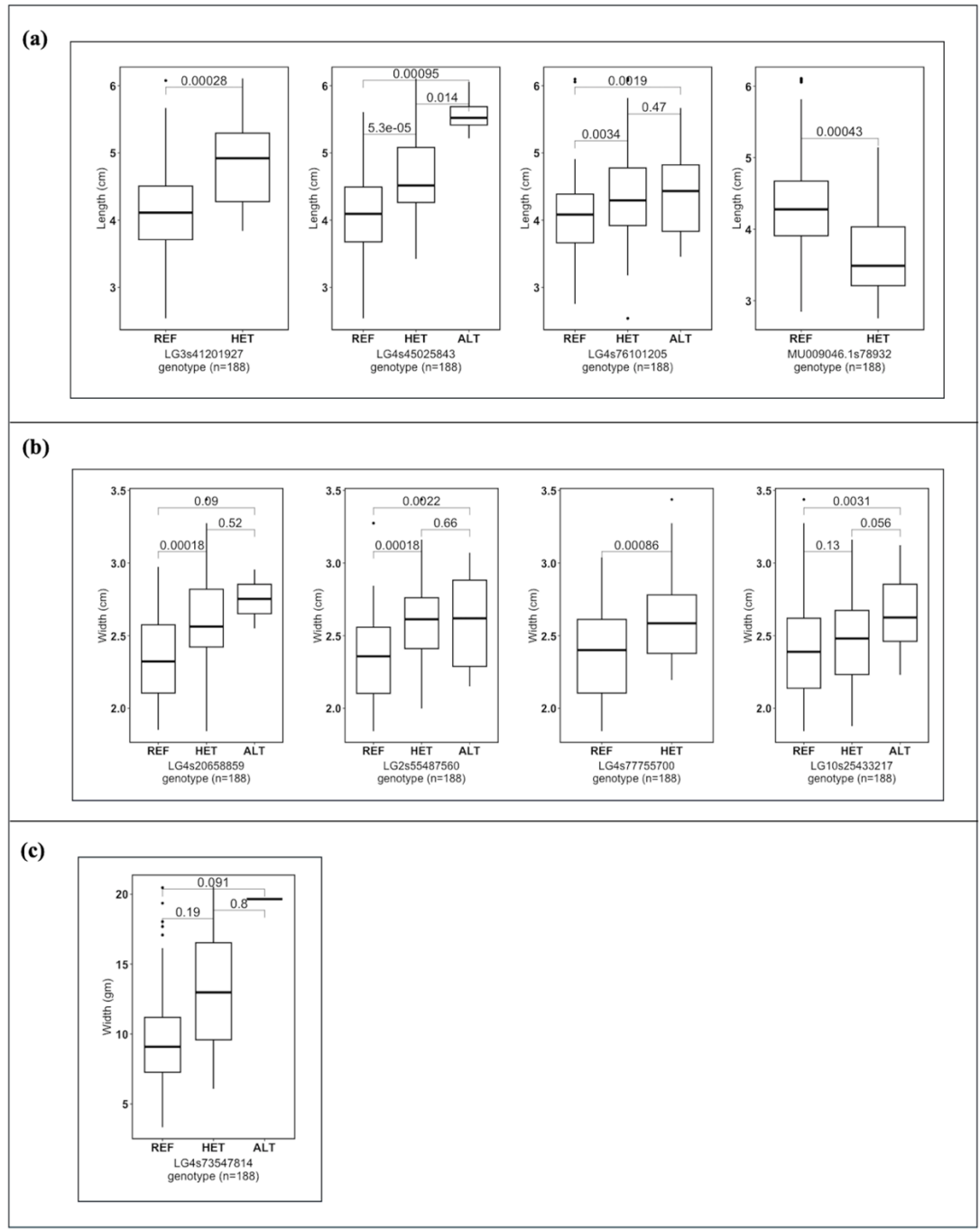

**Figure 3**: Boxplot distribution analysis of fruit size phenotypes (length, width, and weight) by genotypes of significant SNPs from this Genome-wide Association Study (GWAS). p-Values were calculated using the Wilcoxon statistical test. The X-axis represents the SNP genotypes, and the Y-axis represents the R/B phenotypic value. (**a, b, c**) Phenotype-by-genotype analysis of the length, width, and weight phenotypes, respectively.

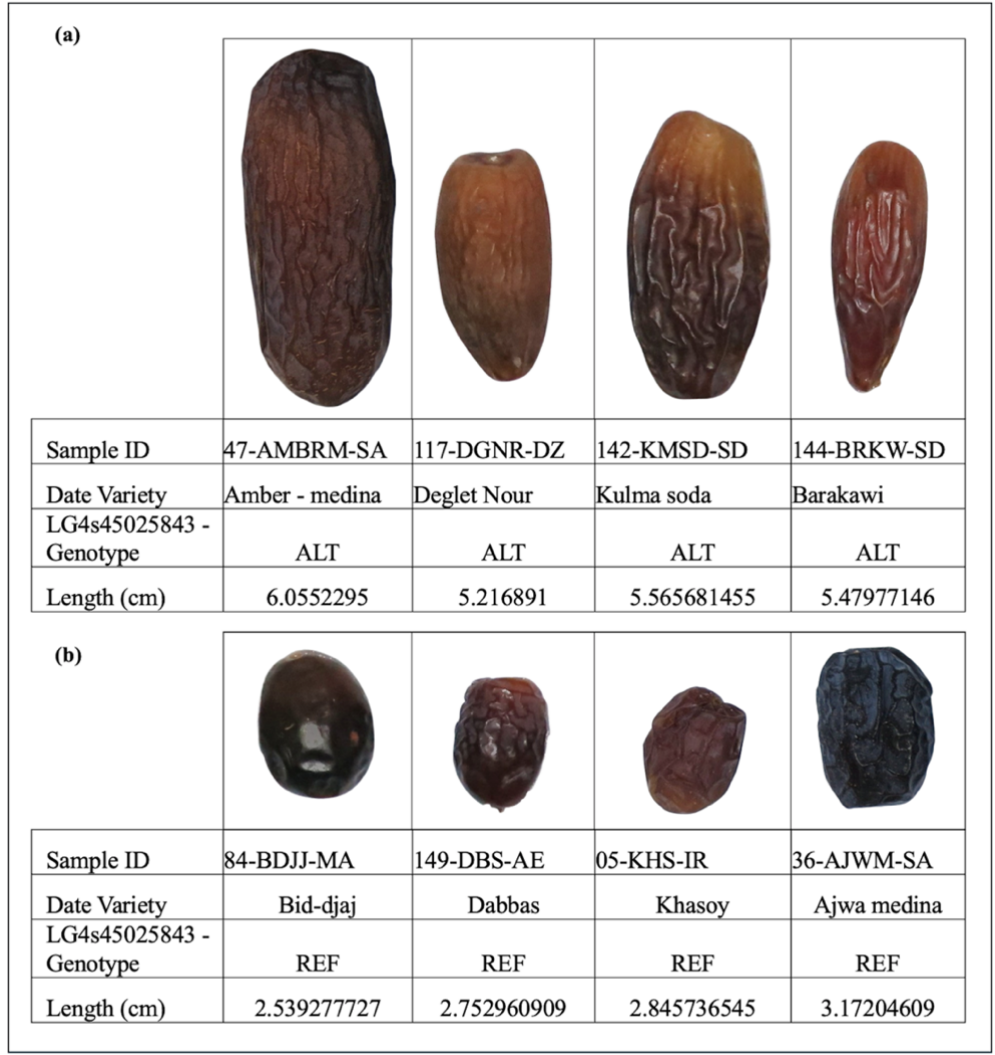

**Figure 4:** Fruit length differences were observed when samples were categorised by the genotypes of SNP LG4s45025843. Fruits were grouped based on whether they were homozygous for the (a) ALT or (b) REF allele of SNP LG18s9876335. The results indicate that the cultivar produces longer fruit when the sample is homozygous for the alternate allele (ALT), while smaller fruit is produced when the sample is homozygous for the reference allele (REF).

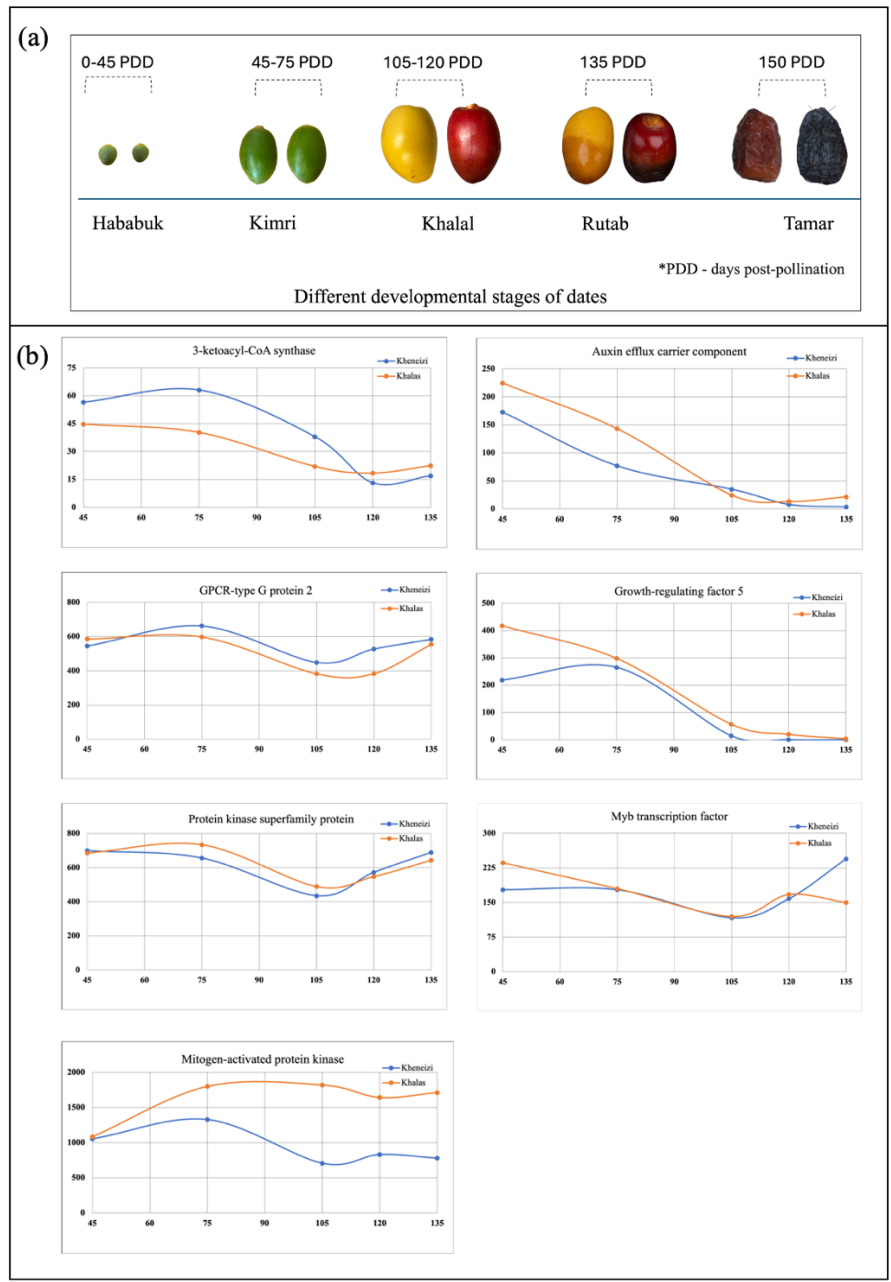

**Figure 5:** RNA-seq analysis of genes identified in the potential regions surrounding GWAS-identified SNPs. Gene expression was assessed across three or four replicates at various fruit development stages for two varieties: Kheneizi (red) and Khalas (yellow). (**a**): Different growth stages of dates. (**b**): Gene expression analysis across Different developmental stages of dates. The X-axis represents the development stage (post-pollination date), while the Y-axis shows the mean normalised expression read count from three or more replicates.

Table 1: List of significant SNPs from GWAS result

| SNP | LG | P.value | MAF | FDR P.Value | Phenotype |
|---|---|---|---|---|---|
| LG1s44795228 | LG1 | 3.47E-13 | 0.19 | 1.23E-06 | Length |
| LG3s41201927 | LG3 | 1.80E-12 | 0.06 | 3.19E-06 | Length |
| LG4s45025843 | LG4 | 9.94E-11 | 0.17 | 0.00011732 | Length |
| LG4s76101205 | LG4 | 1.63E-08 | 0.42 | 0.01444956 | Length |
| MU009046.1s78932 | MU009046 | 2.39E-08 | 0.07 | 0.01693032 | Length |
| LG4s20658859 | LG4 | 3.29E-12 | 0.19 | 1.17E-05 | Width |
| LG2s55487560 | LG2 | 9.70E-10 | 0.29 | 0.00171814 | Width |
| LG4s77755700 | LG4 | 1.02E-08 | 0.16 | 0.01205611 | Width |
| LG10s25433217 | LG10 | 6.06E-08 | 0.24 | 0.0536407 | Width |
| LG4s73547814 | LG4 | 4.66E-13 | 0.09 | 1.65E-06 | Weight |

**Table 2 :** List of functionally significant Gene

| GWAS tag SNP | Phenotype | Gene name | Gene distance form tag SNP | Protein function |
|---|---|---|---|---|
| LG1s44795228 | Length | Protein kinase superfamily protein | 6.5Kb | response to abscisic acid; root epidermal cell differentiation |
| LG3s41201927 | Length | Myb transcription factor | 14.6Kb | abscisic acid-activated signaling pathway; positive regulation of auxin mediated signaling pathway |
| LG4s45025843 | Length | Mitogen-activated protein kinase | 54.4Kb | response to ethylene; auxin-activated signaling pathway; abscisic acid-activated signaling pathway, jasmonic acid mediated signaling pathway; regulation of root meristem growth; regulation of unidimensional cell growth |
| LG4s76101205 | Length | 3-ketoacyl-CoA synthase | 49.4Kb | unidimensional cell growth; |
| MU009046.1s78932 | Length | Auxin efflux carrier component | 32.5Kb | auxin-activated signaling pathway |
| LG2s55487560 | Width | Growth-regulating factor 5 | 73.6Kb | developmental process |
| LG2s55487560 | Width | GPCR-type G protein2 | 65.5 kb | response to abscisic acid; abscisic acid binding; |
| LG4s77755700 | Width | Cytochrome P450, putative | 12.3Kb | positive regulation of cell population proliferation; :regulation of growth rate; positive regulation of organ growth; seed development; |

**Data archiving statement**

Table 1 presents the SNP association results for the size phenotypes of date palm fruit mapped to the reference genome. The gene annotations for the associated linkage groups are provided in Table 2

**Author contributions**

SY designed the study, analyzed the data, conducted bioinformatics analysis and visualisation, and wrote the manuscript. LSM performed the genome sequencing. YAM directed library construction and sequencing. JAM envisioned the project and reviewed the manuscript. KS contributed to the study design and reviewed the manuscript. KFXM also contributed to the study design and reviewed the manuscript.

**Conflicts of interest**

On behalf of all authors, the corresponding author states that there is no conflict of interest.

**Compliance with Ethical Standards**

Human and animal subjects were not included in this study. The authors declare not competing interests.

**Acknowledgements**

We thank Robert Krueger of the USDA, Sean Lahmeyer of Huntington Gardens and Diego Rivera of the University of Murcia for contributing date palm, Phoenix and palm species.

**Funding**

This study was made possible by grant NPRP-EP X-014-4-001 from the Qatar National Research Fund (a member of Qatar Foundation). The funding agency did not participate in the study design, sample collection, analysis, data interpretation or writing of this research.

# 6. Reference

1. Nixon, R.W., *The date palm—"Tree of Life" in the subtropical deserts.* Economic Botany, 1951. **5**(3): p. 274-301.
2. Al-Shahib, W. and R.J. Marshall, *The fruit of the date palm: its possible use as the best food for the future?* Int J Food Sci Nutr, 2003. **54**(4): p. 247-59.
3. Ghnimi, S., et al., *Date fruit (Phoenix dactylifera L.): An underutilized food seeking industrial valorization.* NFS Journal, 2017. **6**: p. 1-10.
4. Awad, M.A., *Increasing the rate of ripening of date palm fruit (Phoenix dactylifera L.) cv. Helali by preharvest and postharvest treatments.* Postharvest Biology and Technology, 2007. **43**(1): p. 121-127.
5. Younuskunju, S., et al., *The genetics of fruit skin separation in date palm.* BMC Plant Biology, 2024. **24**(1): p. 1050.
6. Younuskunju, S., et al., *Genome-wide association of dry (Tamar) date palm fruit color.* The Plant Genome, 2023. **16**(4): p. e20373.
7. Hussain, M.I., M. Farooq, and Q.A. Syed, *Nutritional and biological characteristics of the date palm fruit (Phoenix dactylifera L.) – A review.* Food Bioscience, 2020. **34**: p. 100509.
8. Zaid, A. and E.d.J. Arias-Jimenez, *Date palm cultivation.* FAO Plant Production and Protection Paper (FAO), 1999.
9. Mohamoud, Y.A., et al., *Novel subpopulations in date palm (Phoenix dactylifera) identified by population-wide organellar genome sequencing.* BMC genomics, 2019. **20**(1): p. 1-7.
10. Malek, J.A., et al., *Deletion of beta-fructofuranosidase (invertase) genes is associated with sucrose content in Date Palm fruit.* Plant Direct, 2020. **4**(5): p. e00214.
11. Torres, M.F., et al., *Evidence of Recombination Suppression Blocks on the Y Chromosome of Date Palm (Phoenix dactylifera).* Frontiers in Plant Science, 2021. **12**.
12. Hazzouri, K.M., et al., *Genome-wide association mapping of date palm fruit traits.* Nature Communications, 2019. **10**(1): p. 4680.
13. Al-Dous, E.K., et al., *De novo genome sequencing and comparative genomics of date palm (Phoenix dactylifera).* Nature Biotechnology, 2011. **29**(6): p. 521-527.
14. Mathew, L.S., et al., *A first genetic map of date palm (Phoenix dactylifera) reveals long-range genome structure conservation in the palms.* BMC Genomics, 2014. **15**(1): p. 285.
15. Gillaspy, G., H. Ben-David, and W. Gruissem, *Fruits: A Developmental Perspective.* Plant Cell, 1993. **5**(10): p. 1439-1451.
16. Okello, R.C.O., et al., *What drives fruit growth?* Funct Plant Biol, 2015. **42**(9): p. 817-827.
17. Barry, C.S. and J.J. Giovannoni, *Ethylene and Fruit Ripening.* Journal of Plant Growth Regulation, 2007. **26**(2): p. 143-159.
18. Jahed, K.R. and P.M. Hirst, *Fruit growth and development in apple: a molecular, genomics and epigenetics perspective.* Front Plant Sci, 2023. **14**: p. 1122397.
19. Katel, S., et al., *Impacts of plant growth regulators in strawberry plant: A review.* Heliyon, 2022. **8**(12): p. e11959.
20. Pesaresi, P., et al., *Genetic regulation and structural changes during tomato fruit development and ripening.* Front Plant Sci, 2014. **5**: p. 124.

21. Quinet, M., et al., *Tomato Fruit Development and Metabolism.* Front Plant Sci, 2019. **10**: p. 1554.
22. Sierra-Orozco, E., et al., *Identification and characterization of GLOBE, a major gene controlling fruit shape and impacting fruit size and marketability in tomato.* Horticulture Research, 2021. **8**(1): p. 138.
23. Penchovsky, R. and D. Kaloudas, *Molecular factors affecting tomato fruit size.* Plant Gene, 2023. **33**: p. 100395.
24. Azzi, L., et al., *Fruit growth-related genes in tomato.* Journal of Experimental Botany, 2015. **66**(4): p. 1075-1086.
25. Tanksley, S.D., *The Genetic, Developmental, and Molecular Bases of Fruit Size and Shape Variation in Tomato.* The Plant Cell, 2004. **16**(suppl_1): p. S181-S189.
26. Wang, S., et al., *Arabidopsis ovate family proteins, a novel transcriptional repressor family, control multiple aspects of plant growth and development.* PLoS One, 2011. **6**(8): p. e23896.
27. Sosa, J.M., et al., *Development and application of MIPAR™: a novel software package for two- and three-dimensional microstructural characterization.* Integrating Materials and Manufacturing Innovation, 2014. **3**(1): p. 123-140.
28. Liu, X., et al., *Iterative Usage of Fixed and Random Effect Models for Powerful and Efficient Genome-Wide Association Studies.* PLoS Genet, 2016. **12**(2): p. e1005767.
29. Purcell, S., et al., *PLINK: a tool set for whole-genome association and population-based linkage analyses.* Am J Hum Genet, 2007. **81**(3): p. 559-75.
30. Yin, L., et al., *rMVP: A Memory-efficient, Visualization-enhanced, and Parallel-accelerated Tool for Genome-wide Association Study.* Genomics, Proteomics & Bioinformatics, 2021. **19**(4): p. 619-628.
31. Conesa, A. and S. Götz, *Blast2GO: A comprehensive suite for functional analysis in plant genomics.* Int J Plant Genomics, 2008. **2008**: p. 619832.
32. Cingolani, P., et al., *A program for annotating and predicting the effects of single nucleotide polymorphisms, SnpEff: SNPs in the genome of Drosophila melanogaster strain w1118; iso-2; iso-3.* Fly (Austin), 2012. **6**(2): p. 80-92.
33. Wright, S.I., et al., *The effects of artificial selection on the maize genome.* Science, 2005. **308**(5726): p. 1310-4.
34. Weiss, E., D. Zohary, and M. Hopf, *Domestication of Plants in the Old World - The Origin and Spread of Domesticated Plants in South-west Asia, Europe, and the Mediterranean Basin.* 2012.
35. Allaby, R.G., *Domestication Syndrome in Plants*, in *Encyclopedia of Global Archaeology*, C. Smith, Editor. 2014, Springer New York: New York, NY. p. 2182-2184.
36. Schreiber, M., N. Stein, and M. Mascher, *Genomic approaches for studying crop evolution.* Genome Biology, 2018. **19**(1): p. 140.
37. Meyer, R.S., A.E. DuVal, and H.R. Jensen, *Patterns and processes in crop domestication: an historical review and quantitative analysis of 203 global food crops.* New Phytologist, 2012. **196**(1): p. 29-48.
38. Bons, H. and M. Sidhu, *Role of plant growth regulators in improving fruit set, quality and yield of fruit crops: a review.* The Journal of Horticultural Science and Biotechnology, 2019. **95**: p. 1-10.

39. Zahid, G., et al., *An overview and recent progress of plant growth regulators (PGRs) in the mitigation of abiotic stresses in fruits: A review.* Scientia Horticulturae, 2023. **309**: p. 111621.
40. Alexander, L. and D. Grierson, *Ethylene biosynthesis and action in tomato: a model for climacteric fruit ripening.* Journal of Experimental Botany, 2002. **53**(377): p. 2039-2055.
41. Liu, X., et al., *Cucumber Fruit Size and Shape Variations Explored from the Aspects of Morphology, Histology, and Endogenous Hormones.* Plants (Basel), 2020. **9**(6).
42. Nobusawa, T., et al., *Synthesis of Very-Long-Chain Fatty Acids in the Epidermis Controls Plant Organ Growth by Restricting Cell Proliferation.* PLOS Biology, 2013. **11**(4): p. e1001531.
43. Batsale, M., et al., *Biosynthesis and Functions of Very-Long-Chain Fatty Acids in the Responses of Plants to Abiotic and Biotic Stresses.* Cells, 2021. **10**(6): p. 1284.
44. Kim, J., S.B. Lee, and M.C. Suh, *Arabidopsis 3-ketoacyl-CoA synthase 4 is essential for root and pollen tube growth.* Journal of Plant Biology, 2021. **64**(2): p. 155-165.
45. Horiguchi, G., G.T. Kim, and H. Tsukaya, *The transcription factor AtGRF5 and the transcription coactivator AN3 regulate cell proliferation in leaf primordia of Arabidopsis thaliana.* The Plant Journal, 2005. **43**(1): p. 68-78.
46. Wu, W., et al., *Growth-regulating factor 5 (GRF5)-mediated gene regulatory network promotes leaf growth and expansion in poplar.* New Phytologist, 2021. **230**(2): p. 612-628.
47. Vercruyssen, L., et al., *GROWTH REGULATING FACTOR5 stimulates Arabidopsis chloroplast division, photosynthesis, and leaf longevity.* Plant Physiol, 2015. **167**(3): p. 817-32.
48. Kim, J.H., *Biological roles and an evolutionary sketch of the GRF-GIF transcriptional complex in plants.* BMB reports, 2019. **52**(4): p. 227.
49. Okal, M., et al., *Auxin polar transport and flower formation in Arabidopsis thaliana transformed with indoleacetamide hydrolase (iaaH) gene.* Plant and cell physiology, 1999. **40**(2): p. 231-237.
50. Mohan, A., et al., *Molecular Characterization of Auxin Efflux Carrier- ABCB1 in hexaploid wheat.* Scientific Reports, 2019. **9**(1): p. 17327.
51. Leyser, O., *Plant hormones: Ins and outs of auxin transport.* Current Biology, 1999. **9**(1): p. R8-R10.
52. Luschnig, C., *Auxin transport: Why plants like to think BIG.* Current Biology, 2001. **11**(20): p. R831-R833.
53. Benková, E., et al., *Local, Efflux-Dependent Auxin Gradients as a Common Module for Plant Organ Formation.* Cell, 2003. **115**(5): p. 591-602.
54. Wang, Y., et al., *Auxin efflux carrier ZmPIN1a modulates auxin reallocation involved in nitrate-mediated root formation.* BMC Plant Biology, 2023. **23**(1): p. 74.
55. Swarup, R. and R. Bhosale, *Developmental roles of AUX1/LAX auxin influx carriers in plants.* Frontiers in Plant Science, 2019. **10**: p. 1306.
56. Siddiq, M. and I. Greiby, *Overview of Date Fruit Production, Postharvest handling, Processing, and Nutrition*, in *Dates*. 2013. p. 1-28.